\documentclass[preprint,showpacs,preprintnumbers,amsmath,amssymb]{revtex4}

\usepackage{graphicx}

\usepackage{dcolumn}

\usepackage{bm}

\begin{document}

\title{Sales Distribution of Consumer Electronics}

\author{Ryohei Hisano$^{a}$\;\footnote{Corresponding author: Ryohei Hisano
\newline Address: 2-1, Naka, Kunitachi City, Tokyo 186-8603, Japan
\newline Tel: +81-42-580-8357 Fax: +81-42-580-8357
\newline E-mail address: em072010@yahoo.co.jp
}, Takayuki Mizuno$^{b}$}%
\address{$^{a}$Graduate School of Economics, Hitotsubashi University, 2-1 Naka, Kunitachi City, Tokyo 186-8603, Japan 
\linebreak $^{b}$Institute of Economic Research, Hitotsubashi University, 2-1 Naka, Kunitachi City, Tokyo 186-8603, Japan 
}
\date{August, 30 2010}

\begin{abstract}
Using the uniform most powerful unbiased test, we observed the sales distribution of consumer electronics in Japan on a daily basis and report that it follows both a lognormal distribution and a power-law distribution and depends on the state of the market. We show that these switches occur quite often. The underlying sales dynamics found between both periods nicely matched a multiplicative process. However, even though the multiplicative term in the process displays a size-dependent relationship when a steady lognormal distribution holds, it shows a size-independent relationship when the power-law distribution holds. This difference in the underlying dynamics is responsible for the difference in the two observed distributions.

\

\end{abstract}


\maketitle

\section{Introduction}

Since Pareto pointed out in 1896 that the distribution of income exhibits a heavy-tailed structure [1], many papers has argued that such distributions can be found in a wide range of empirical data that describe not only economic phenomena but also biological, physical, ecological, sociological, and various man-made phenomena [2]. The list of the measurements of quantities whose distributions have been conjectured to obey such distributions includes firm sizes [3], city populations [4], frequency of unique words in a given novel [5-6], the biological genera by number of species [7], scientists by number of published papers [8], web files transmitted over the internet [9], book sales [10], and product market shares [11]. Along with these reports the argument over the exact distribution, whether these heavy-tailed distributions obey a lognormal distribution or a power-law distribution, has been repeated over many years as well [2]. In this paper we use the statistical techniques developed in this literature to clarify the sales distribution of consumer electronics.

To illustrate the heavy-tailed distribution's appearance, random growth processes are widely used as the approximation of its underlying dynamics. Gibrat, who built upon Kapteyn and Uven's work, was the first to propose the simplest form of this type of model, usually known as the multiplicative process, to describe the appearance of heavy-tailed distributions in firm size distributions [12]. His work is significant in market structure literature [13]. Even 70 years after Gibrat's book, more and more measures of quantities are found that are conjectured to obey this type of process.

Among recent works, Fu et al. [14] is crucial because it was perhaps the first work to consider the hierarchical structure of institutions. No one denies that firms grow in size and scope and that such growth is heavily influenced by the successful launch of new products. Fu et al. modeled products as elementary sales units assuming that they evolve based on a random multiplicative process. They extended the usual model of proportional growth to illustrate the size variance relationship found in growth distributions at different levels of aggregation in the economy by considering hierarchical structure. 
	
Many studies have analyzed product sales. Sornette et al. [10] used a book sales database from Amazon.com and performed a time series analysis of book sales by classifying endogenous and exogenous shocks. With a database of newspaper and magazine circulation, Picolli et al. [15] used the multiplicative process to illustrate the link between tent-shaped log growth distributions and the power-law distributions found in the growth of newspaper and magazine sales. However the exact dynamics of product sales remains an open question.

In this paper we clarify the distribution and the dynamics of the sales of consumer electronics using a unique database of product sales from the Japanese consumer electronics market. The data were recorded daily, making it possible to track the actual sales volume of each product in a more detailed manner and to model the dynamics from a more empirically driven approach. We numerically analyzed more than 1200 sales distributions recorded on a daily basis from October 1 2004 to February 29 2008. Using the uniform most powerful test, we statistically show that sale distributions differ among different periods and occasionally exhibit a power-law behavior. We also show with the multiplicative process that the underlying ingredients of stochastic growth itself are different among these periods. Moreover our findings are compatible with the mathematical results reported by Ishikawa et al. [16].

The paper is organized as follows. Section 2 provides an overview of our data set. Section 3 introduces the sales distribution of consumer electronics, and Section 4 illustrates our statistical technique regarding the verification of a true power-law distribution. Using this statistical technique, in Section 5, we show why the power-law behavior found in Section 3 can be considered a genuine power-law behavior. Section 6 reports how sales distribution changes over time. Sales distribution exhibits both power-law and lognormal distributions. Section 7 focuses on the underlying dynamics of sales, providing another source of evidence that the dynamics of sales differs among different periods. Section 8 provides further discussion and a conclusion.

\section{Sales data of consumer electronics}
Consumer electronics chains sell products such as TVs, personal computers, audio devices, refrigerators, digital cameras, air conditioners, and DVD recorders. Their annual revenue amounts to 5.9 trillion yen in Japan. In this paper we investigate distribution using the sales data of digital cameras from 23 different consumer electronics chains in Japan collected by a private company called BCN Inc. This dataset covers about 45{\%} of all consumer electronics chains in Japan including over 1,400 retail stores [17]. The data were recorded daily covering the period from October 1 2004 to April 30 2008.

\section{Sales distribution of consumer electronics}
We focused on the top selling products using cumulative distribution $P_>(S)$ defined as
\begin{equation}
P_>(x):=Pr[X \ge x]= \int_x^\infty f(x')dx'
\end{equation}
where $f(x)$ describes the probability density function. The cumulative distribution of the sales volume of digital cameras on April 1 2005 is shown on a double logarithmic scale (Fig. 1). It exhibits a heavy-tailed structure. To investigate the exact characteristic of its distribution, we also depict the maximum likelihood estimate of a lognormal distribution, assuming that all values above 1 obeyed a lognormal distribution and the maximum likelihood estimate of a power-law distribution, assuming that all values above 16 obeyed a power-law distribution. A lognormal distribution fits nicely for almost all points except the last three. For the points above 16 including these last three points, at first glance it seems that a power-law distribution fits better. In this paper we numerically judge whether a simple lognormal distribution or a lognormal distribution with a power-law tail displays a better fit using the statistical technique developed by Malevergne et al. [18]. The importance of distinguishing between these two distributions lies in the fact that not only does the tail describe the top selling products but these products which seems to exist in the power-law region account for about 80{\%} of total sales; identifying the dynamics of these top selling products is important.

\section{Testing a power-law distribution at the tail}
To judge whether a power-law distribution or a lognormal distribution displays a better fit for values over a threshold, one natural way is using a model selection technique between a power-law distribution and a singly truncated lognormal distribution that puts the truncation point identically as the lower bound of a power-law distribution (for instance, see Clauset et al. [19]). The basic change of the variables shows that a logarithm of a random variable, which obeys a power-law distribution, is an exponential distribution, but a logarithm of a singly truncated lognormal distribution is a singly truncated normal distribution. Hence the test of a power law against a singly truncated lognormal is equivalent to testing an exponential distribution against a singly truncated normal distribution in the log-size distribution of the original measure of quantity. 

Next, as shown by Castillo [20], an exponential distribution and a singly truncated normal distribution have the following relationship:

\begin{equation}
f_{STN}(x;\alpha,\beta,A)\to f_{exp}(x;\lambda)1_{x_>A}\;\;\;\;\;as\;\;\;(\alpha,\beta)\to(\lambda,0),
\end{equation}
where $A$  denotes the truncation point and $\alpha$  and $\beta$ are the parameters of a singly truncated normal distribution with the following relationship:

\begin{equation}
\alpha:=-\frac{\mu-A}{\sigma^2}\;\;\beta:=\frac{1}{2\sigma^2}
\end{equation}
where $\mu$ is the usual mean, $\sigma$ is the standard deviation. This implies that an exponential distribution is in the boundary line of a singly truncated normal distribution. This relationship illustrates why a singly truncated normal distribution (singly truncated lognormal distribution) so closely resembles an exponential distribution (power-law distribution) if $\beta$ becomes incresaingly close to 0. Fig. 2 shows the maximum likelihood estimate assuming an exponential distribution and a singly truncated normal distribution for the log-size distribution of digital camera sales on April 1 2005, setting the truncation point as $A=log(16)$. Observe from the maximum likelihood estimate that a singly truncated normal distribution with sufficiently small   closely resembles an exponential distribution. 
	
	Considering this relationship, a natural test to distinguish an exponential distribution from a singly truncated normal alternative is to test the departure from the exponential form (null hypothesis) against the singly truncated normal alternative (alternative hypothesis) using the likelihood ratio test that evaluates statistic 

\begin{equation}
W=2(L(\hat{\theta})-L(\tilde{\theta}))
\end{equation}
where $L$ denotes log-likelihood function $\hat{\theta}=(\hat{\alpha},\hat{\beta})$, which is the maximum likelihood estimate under the full model, and $\tilde{\theta}=(\tilde{\lambda},0)=(\frac{1}{\bar{x}},0)$ in its exponential form. Castillo and Puig [21] showed the following: 1) the likelihood ratio test is the uniform most powerful unbiased test in this case; 2) the likelihood ratio test could easily be performed using the clipped coefficient of variation (i.e. $c=min\{1,\bar{c}\}$, where $\bar{c}$ is the empirical coefficient of variation); and 3) the critical region of the test could be approximated with a high degree of accuracy even for small samples using saddle point approximation. Malevergne et al. [18], who discussed whether a lognormal suffices or a power-law distribution shows a better fit for the upper tail of the size distribution of US city size data, concluded that the upper tail of the size distribution of US cities is in fact a power law.

Figure 3 shows the test using the sales distribution found in April 1 2005. Starting from the top 10 products we recursively calculated the p-value of the test using Castillo and Puig's method. We then calculated the point where the p-value of the test first falls within the critical region (in this paper, $\alpha=0.1$) minus 1. For the sales distribution found in April 1 2005 this point is 68. This implies that for the 68 points above this threshold the power-law distribution is not rejected and shows that the upper tail of the distribution of sales found in April 1 2005 was well fit by a power-law distribution. 

\section{Distribution analysis of sales}
	In a small sample data set, we often observe "power-law behavior" (straight line in the cumulative distribution depicted on a double logarithmic scale) even if it were actually sampled from a theoretical lognormal distribution. Fig. 4 shows two cases that plot the cumulative distribution of synthetic data sets randomly sampled from a theoretical lognormal distribution with the same parameters as the sales distribution of digital cameras on April 1 2005 (Fig. 1). In the one case depicted in the left panel, note that the tail follows a lognormal distribution. However, in the other, even if we used the statistical technique explained in Section 4, the lower bound estimated from the procedure returns a value of 77 for the distribution denoted in circles, confirming a power-law behavior at the tail.
	
	Hence to judge whether distributions found in a certain period are well described by a power-law distribution we must consider all the distributions found during that period. The left panel of Fig. 5 shows the estimated lower bound from the 107 dates during January 22 2005 to May 8 2005 and the right panel shows the estimated lower bound of the first 107 synthetic data sets randomly sampling from a theoretical lognormal distribution with the same parameter as the sales distribution found on April 1 2005. The lower bound from the real data is clearly quite stable, which proves that the power-law behavior found in the sales distribution of April 1 2005 reflected a generating process that produces a genuine power-law behavior and not the result of a process that generates a lognormal distribution. Fig. 6 also shows their probability density, confirming that the behaviors found in Fig. 5 are also quite statistically different. Fig. 7 shows the time evolution of the power-law exponent during January 22 2005 to May 8 2005. The power-law exponent is stable and fluctuates around value $\mu=1.3\pm0.1$ which is quite close to the power law exponent found for city size [18] and wealth [22]. The period when the power-law behavior becomes stable was repeatedly found and shows that the lognormal distribution does not sufficiently describe the sales distribution of digital cameras on a daily basis.

\section{How distribution changes over time}
Next we focus on all the other dates in our data set. Fig. 8 shows their estimated rank thresholds from October 1 2004 to February 29 2008. The period at which we successively observed high estimated rank thresholds is not only January 22 2005 to May 8 2005 but is also found in other parts of the data. However, there is a period when the estimated rank threshold does not behave as if a power-law behavior really exists at its distribution tail: period January 16 2006 to August 8 2006 (Fig. 8). Fig. 9 shows a typical cumulative distribution observed during this period. The lognormal distribution adequately explains the sales distribution for all points. Fig. 10 also shows the histogram of the estimated rank threshold of this period. For this period a simple lognormal distribution adequately describes the distribution of sales. Therefore we conclude that sales distribution reflects when they were observed.
\section{Sales dynamics of consumer electronics}
	It is well known that proportional growth principle applies to firm growth [23].  Fu et al showed that not only does this principle apply to firms, but it also applies to different aggregation of the economy from countries, industry sector, and to products showing theoretically that this stems from its elementary sale unit (i.e. sales of products) evolving accroding to a random multiplicative process [14,24].  Sakai and Watanabe investigated further this issue confirming that dynamics of products determines firms growth by analyzing sales of products sold at grocery stores in Japan [25].  Picoli et al used the multiplicative process to model the dynamics of circulation of newspapers and magazines [15].  Motivated by these literature we would use
\begin{equation}
S(t+1)=|b(t)S(t)+\epsilon(t)|\;\;\;\;\;\;\;\epsilon(t)\sim Gaussian(0,\sigma)
\end{equation}
to describe the underlying dynamics of sales. This assumes a preferential like model for sales which requires age and average sales of products to correlate in a exponential fashion if the distribution of lifetime is exponential.  It is reported that the lifetime distribution often follows exponential functions in competitive markets [26].  Although this relationship could not be easily verified directly with products itself because lifetime of digital camera is short (usually 6 to 12 month) due to product turnover, this could be roughly verified when we observe the average daily market share of brands during there lifetime with their age (fig.11).

	If multiplicative noise $b$ is independent of the former size of $S$, then $S$ leads to a steady power-law distribution [27]. However if it is size dependent, $S$ departs from a power-law distribution [28]. In this section, we show that sales dynamics follow this multiplicative process and use it to reexamine the differences in the sales distribution found in Section 6 from the usually assumed elements of a stochastic growth process.
	
	When BCN Inc. collected this dataset, they made new contracts with other stores to offer sales data and generated an apparent artificial change of product sales along time. To cope with this problem, we introduce normalized sales, $\bar{S}_i(t)=S_i(t)/\frac{1}{n}\sum_{i=1}^{n}S_i(t)$, instead of actual sales to compare two distant periods. Here, n is number of products in the market. All the results in this section can also be reproduced using market share, $\hat{S}_i(t)=S_i(t)/\sum_{i=1}^{n}S_i(t)$, as well. To begin our empirical investigation we cut the scatter plots of both periods into equal logarithmic bins: $0.7\le\bar{S}_{low}<2.175\le\bar{S}_{mid}<6.76\le\bar{S}_{high}<21$  (Fig. 12). The basic idea is to observe whether the distribution of sales growth for one week, $\frac{\bar{S}_i(t+1)}{\bar{S}_i(t)}$, depends on $\bar{S}_i(t)$.
	
	We saw in Section 6 that the tail property of the sales distribution follows a power law for January 22 2005 to May 8 2005 and a lognormal for January 16 2006 to August 8 2006 (Fig. 8). Hence, as shown in Fig. 13, we compare the distribution of log sales growth for 1 week, $log\frac{\bar{S}_i(t+1)}{\bar{S}_i(t)}$, observed during periods January 22 2005 to May 8 2005 and January 16 2006 to August 8 2006. Note that while the positive values of the middle and high areas are quite different during January 16 2006 to August 8 2006, they seem to coincide for the log growth distributions observed during January 22 2005 to May 8 2005. Note also that log growth distribution could be well described as a double exponential distribution and that for the negative log growth rates, the probability density coincides. This suggests that while the multiplicative term for the high and middle areas is size independent during January 22 2005 to May 8 2005, it is size dependent during January 16 2006 to August 8 2006 and shows different behaviors. 
	
	The same observation can also be made using the two sample Kormogorv-Smirnov tests. Table 1 shows the p-value from the test for two pairs, "high vs middle" and "middle vs low", for two periods, January 22 2005 to May 8 2005 and January 16 2006 to August 8, respectively. The only pair for which the test does not reject the null hypothesis is the "high and middle" pair found in January 22 2005 to May 8 2005.

	From these observations, the sales dynamics can be described by the multiplicative process:
\begin{equation}
\bar{S}(t+1)=|b(\bar{S}(t))\bar{S}(t)+\epsilon(t)|\;\;\;\;\;\;\;
b(\bar{S}(t))=
\left\{
\begin{array}{cc}
b_{low}(\bar{S}(t)) & \mbox{if $\bar{S}(t)<2.175$} \\
b_{mid}(\bar{S}(t))=b_{high}(\bar{S}(t))  & \mbox{if $\bar{S}(t) \ge 2.175$}
\end{array}
\right.
\end{equation}
where $t$ describes the time during January 22 2005 to May 8 2005 and
\begin{equation}
\bar{S}(t+1)=|b(\bar{S}(t))\bar{S}(t)+\epsilon(t)|\;\;\;\;\;\;\;
b(\bar{S}(t))=
\left\{
\begin{array}{ccc}
b_{low}(\bar{S}(t)) & \mbox{if $\bar{S}(t)<2.175$} \\
b_{mid}(\bar{S}(t)) & \;\;\;\;\;\;\;\;\;\;\;\mbox{if $2.175 \le \bar{S}(t) <6.76$} \\
b_{high}(\bar{S}(t)) & \!\!\mbox{if $6.76 \le \bar{S}(t)$}
\end{array}
\right.
\end{equation}
where $t$ describes the time during January 16 2006 to August 8 2006. Here, $\epsilon(t)\sim Gaussian(0,\sigma)$. Where $\bar{S}_i(t)$ is large, the log growth distribution displays a size-independent relationship with $\bar{S}_i(t)$ during January 22 2005 to May 8 2005, but it displays a size-dependent relationship with $\bar{S}_i(t)$ during January 16 2006 to August 8 2006. As shown by Ishikawa et al. [16], if the log growth distribution is well described by a double exponential distribution and the probability density coincides for negative log growth rates but exhibits a size dependent relationship for positive values, then the multiplicative process described in Eq. (7) will generate a steady lognormal distribution. Recall that during period January 16 2006 to August 8 2006 this condition is satisfied. On the other hand, as shown by Takayasu et al. [27], Eq. (5) theoretically generates a power-law distribution when the growth distribution is independent of $\bar{S}_i(t)$. Therefore while Eq. (6) generates a distribution with a power-law tail, Eq. (7) generates a simple lognormal distribution that explains the difference in the underlying dynamics for the two periods in which we observed different distributions.
\section{Conclusion and further discussions}
This paper showed how the sales distributions of products evolves when we observed them daily. We showed that the distribution of the top ranking products switches between lognormal and power-law distributions depending on the timing, suggesting that the underlying dynamics among these periods differs. This structural difference in the underlying dynamics was well established from the usually assumed ingredients of the growth process as well providing another source of evidence that the dynamics between these two periods differ. Our result is mathematically compatible with Ishikawa et al. [16], who illustrated the appearance of both power-law and lognormal distributions under a multiplicative process. We only investigated digital cameras in this paper; however such findings can be established with many other products in consumer electronics markets as well.

An interesting question to ponder is why the switch behavior found in Section 6 occurred. In our case the main source of the switch can probably be explained by product turnover.  In product markets such as the digital camera market product life cycle is short taking only about 6 to 12 month for a particular brand to change from an old product to new one due to product competition.  Those product turnover usually takes place on February and August before the aggregate demand for digital cameras starts to rise.  Fig. 14 shows the time evolution of the number of products and the lower bound.  As denoted in fig. 14, the periods coincide when we observed a steady power-law behavior and a rapid increase in the number of products (i.e. when rapid product turnover take place), explaining the switch behavior from a lognormal to a power law.  When a large number of new products are born simultaneously, sales distributions are generated by a mixture of old and new products making sales dynamics to be more accurately described as Gibrat's law (i.e. size independent growth rate).  An empirical study taking these product turnover effect is future work.

Since researchers are equipped with more detailed data from actual markets we can investigate actual market coordination in a more detailed sense. These studies are important not only for economics literature but also for physics because such social systems as the market are one natural laboratory for investigating coordination under complex systems. We hope this line of studies continues to be fruitful for both physics and economics.
\begin{acknowledgments}
	The authors want to thank D. Sornette, A. Ishikawa, and S. Fujimoto for their helpful suggestions concerning this work. Without them this work would never have reached its current level. We would also like to thank BCN Inc. for providing its data. Many thanks goes to the two anonymous referees who gave us helpful comments as well.  This research is a part of a project entitled: Understanding Inflation Dynamics of the Japanese Economy, funded by JSPS Grant-in-Aid for Creative Scientific Research (18GS0101). Takayuki Mizuno was supported by funding from the Kampo Foundation 2009.
\end{acknowledgments}

\newpage

\begin{figure}

\centerline{\includegraphics[width=14cm,bb= 274 132 548 384]{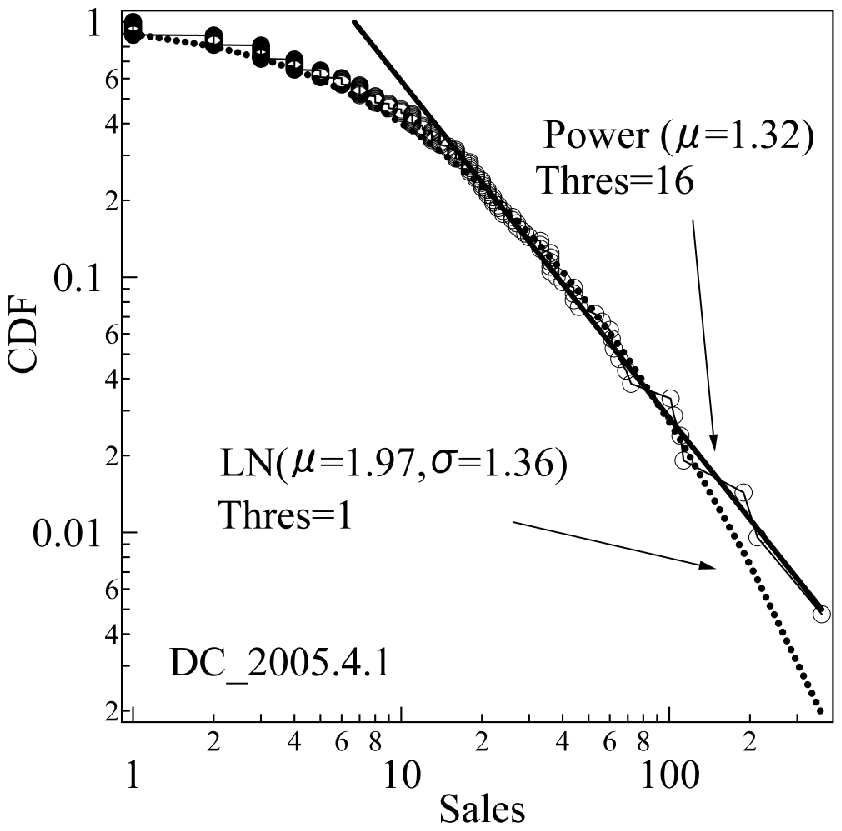}}

\caption{Cumulative distribution of sales volume of digital cameras sold on April 1 2005. Slashed line shows fitted maximum likelihood estimate assuming all points above 1 obeyed a lognormal distribution, and continuous line shows fitted maximum likelihood estimate assuming all points above 16 obeyed a power-law distribution. Parameters of both distributions are depicted as well.}

\end{figure}

\newpage
\begin{figure}

\centerline{\includegraphics[width=14cm,bb=321 198 594 448]{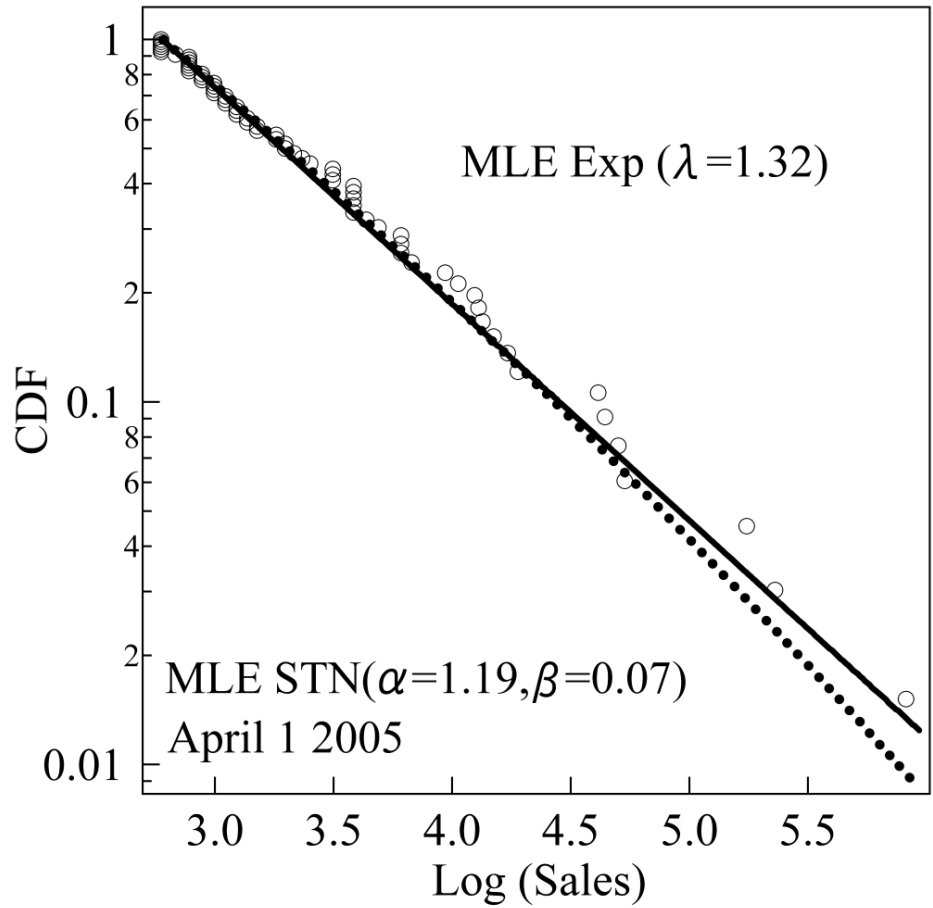}}

\caption{ Log-size distribution of sales distribution found on April 1 2005 for values over $A=log(16)$. Maximum likelihood estimates of both exponential and singly truncated normal distributions are depicted along with their parameters.
}

\end{figure}

\newpage
\begin{figure}

\centerline{\includegraphics[width=19cm,bb=165 212 716 463]{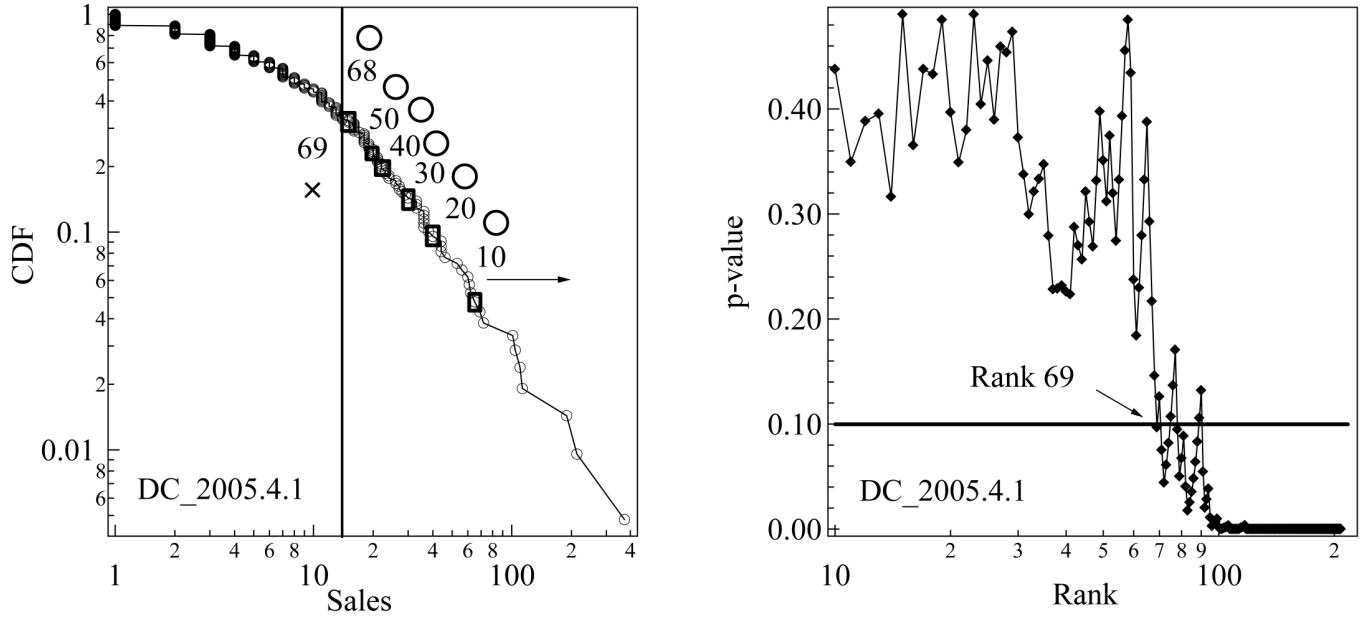}}

\caption{Right panel depicts p-value of test of null hypothesis where distribution's upper tail is power against alternative singly truncated lognormal distribution as a function of rank threshold.
}

\end{figure}

\clearpage
\begin{figure}

\centerline{\includegraphics[width=19cm,bb=162 200 707 451]{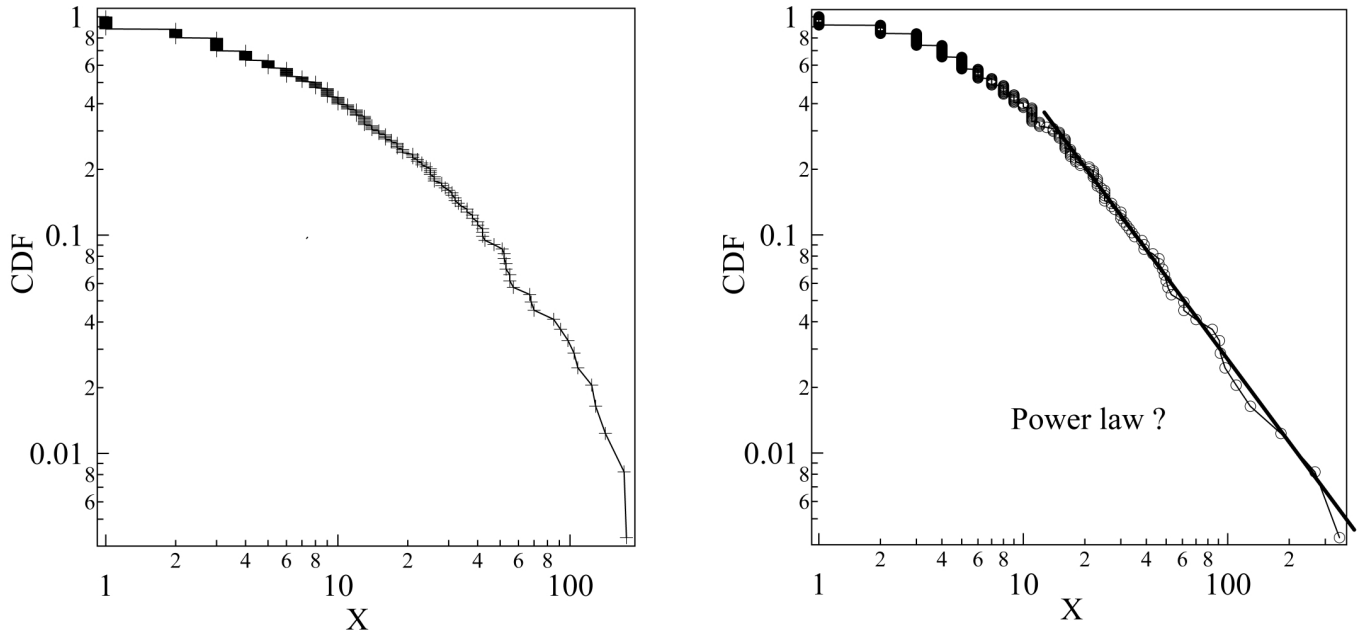}}

\caption{Two examples of randomly sampling from a lognormal distribution with parameters $\mu=1.97, \sigma=1.36$. There are 250 points in both distributions. Note the power-law behavior at the distribution's tail denoted by circles. Estimated lower bound for crosses is 9 and 77 for circles.
}

\end{figure}

\clearpage
\begin{figure}

\centerline{\includegraphics[width=19cm,bb=167 240 705 434]{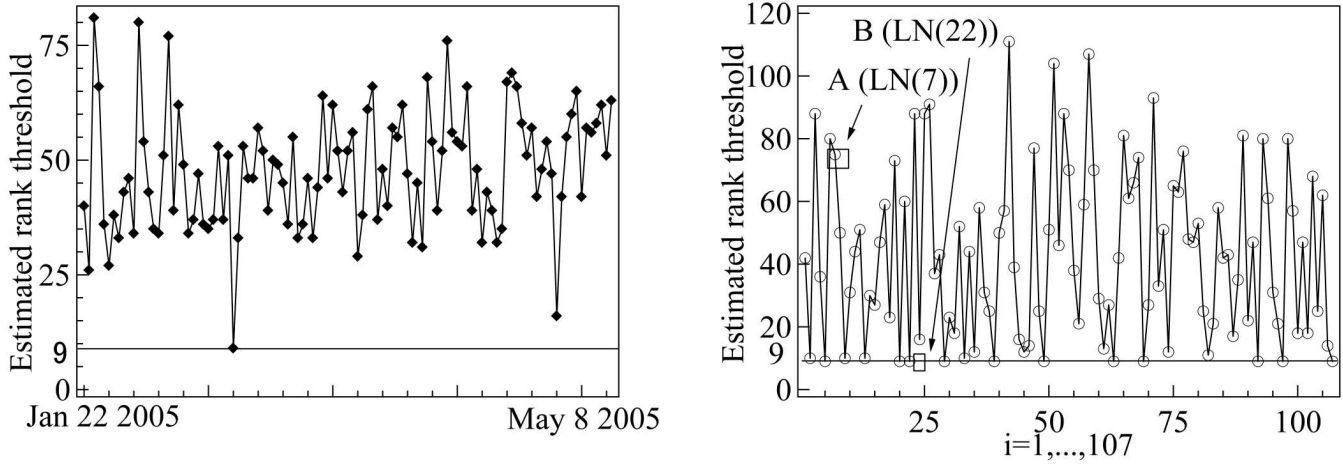}}

\caption{Left panel shows estimated rank threshold from distribution of sales volume from January 22 2005 to May 9 2005. Right panel shows estimated rank threshold for first 107 synthetic data sets $(LN(1),LN(2),...,LN(107))$ obtained in experiment. Estimated rank threshold of 7th and 22nd data sets denoted as A and B are used to depict Fig. 4.
}

\end{figure}

\clearpage
\begin{figure}

\centerline{\includegraphics[width=14cm,bb=277 236 598 459 ]{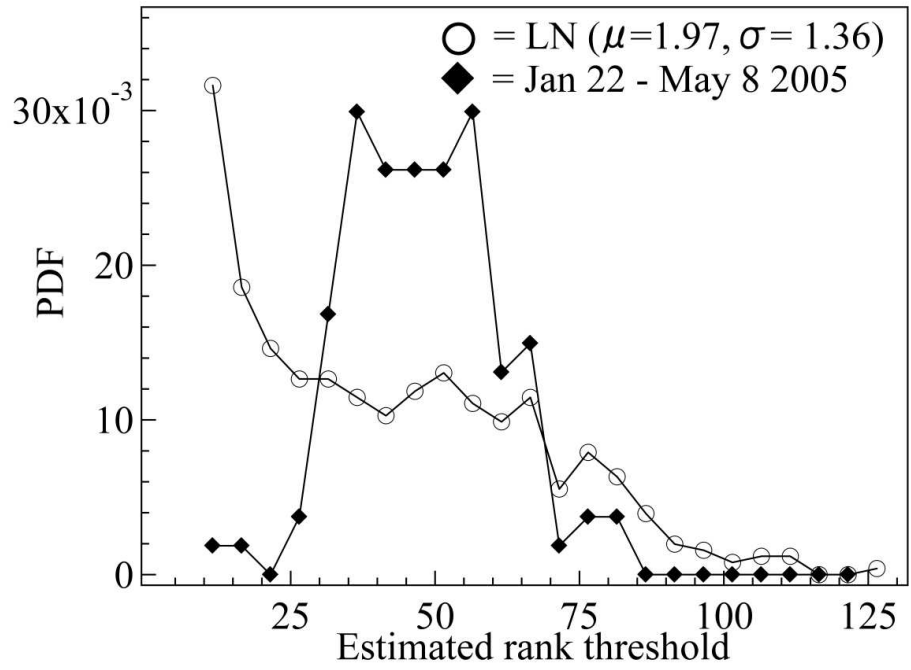}}

\caption{Histogram comparison of estimated rank threshold. Circles denote histogram from synthetic data sets, and diamonds denote histogram from real data.
}

\end{figure}

\clearpage
\begin{figure}

\centerline{\includegraphics[width=14cm,bb=264 145 537 396 ]{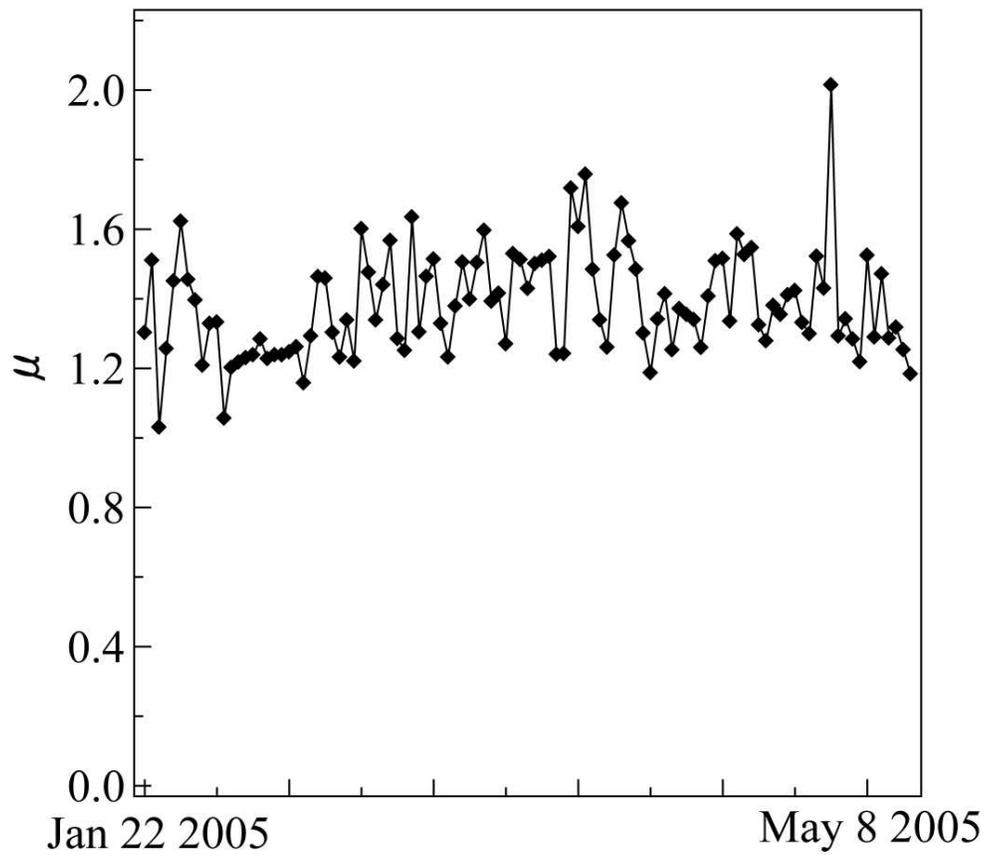}}

\caption{Time evolution of power-law exponent found during period January 22 2005 to May 8 2005.
}

\end{figure}

\clearpage
\begin{figure}

\centerline{\includegraphics[width=14cm,bb=312 234 635 426 ]{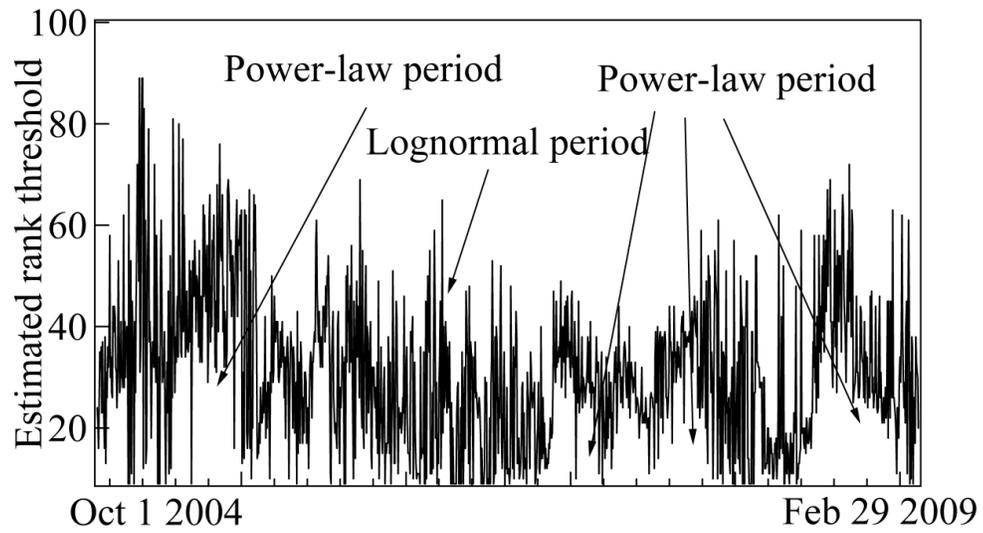}}

\caption{Time evolution of estimated rank threshold for entire period (October 1 2004 to February 29 2008).
}

\end{figure}

\clearpage
\begin{figure}

\centerline{\includegraphics[width=14cm,bb=257 268 530 519 ]{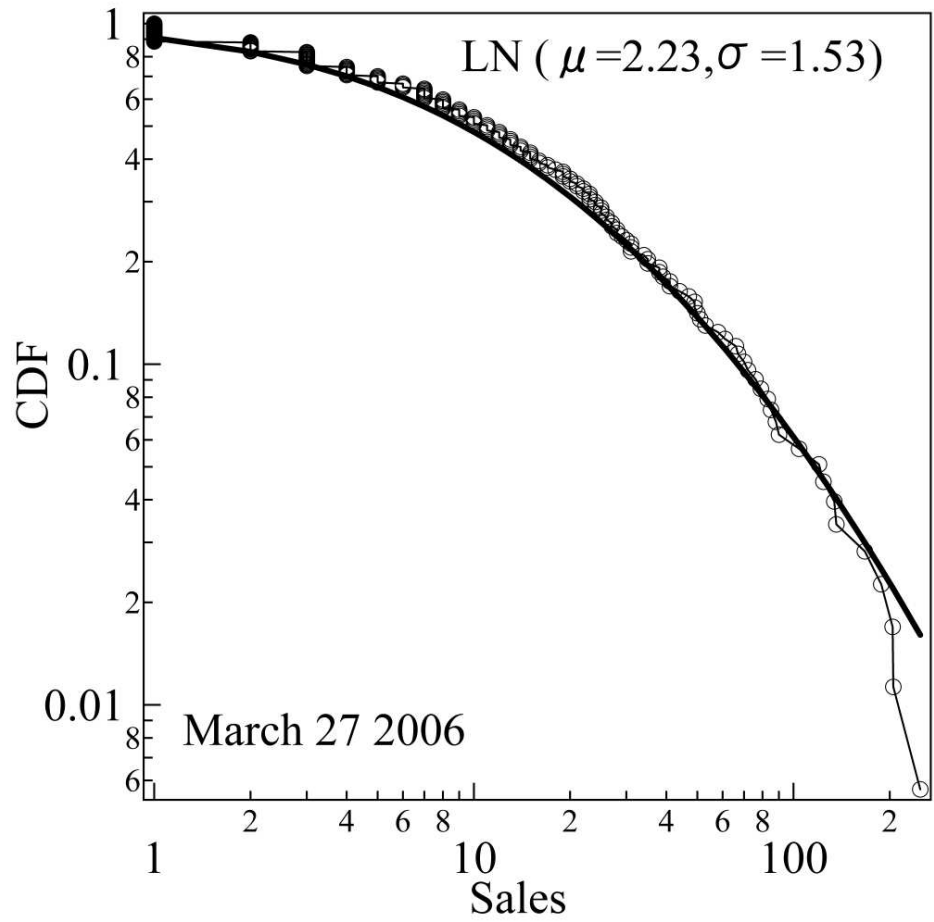}}

\caption{Cumulative distribution of sales volume of digital cameras sold on March 27 2006. Continuous line shows fitted maximum likelihood estimate assuming that all points above 1 obeyed a lognormal distribution. Maximum likelihood estimate of parameters is written inside parentheses.
}

\end{figure}

\clearpage
\begin{figure}

\centerline{\includegraphics[width=14cm,bb=262 171 553 422]{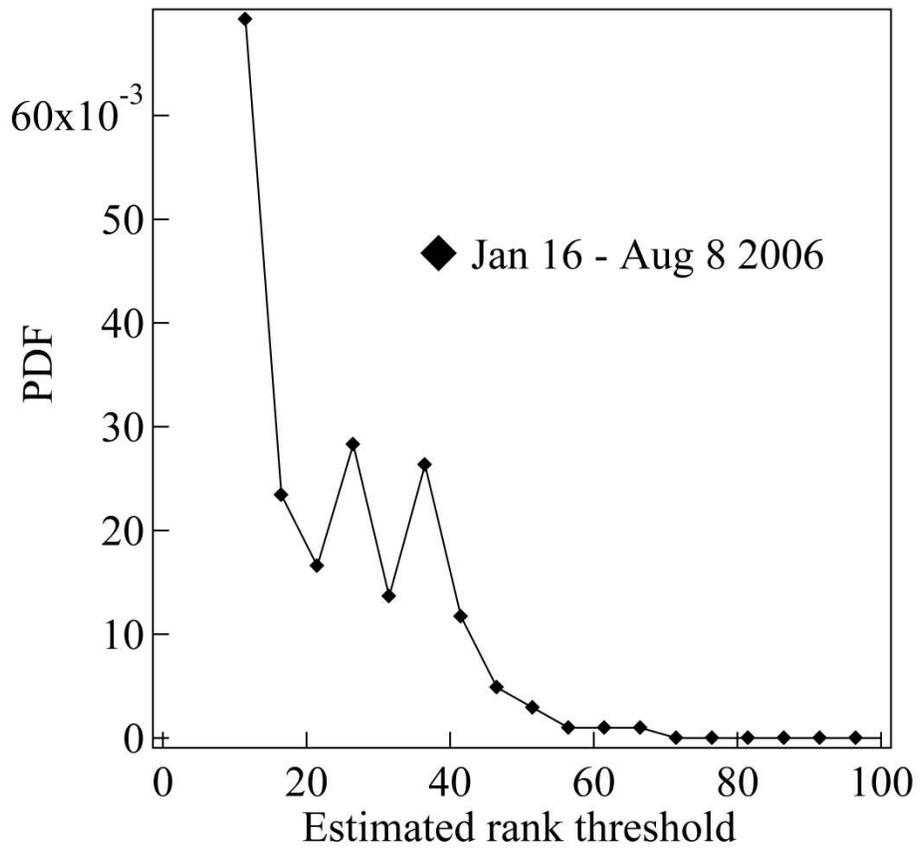}}

\caption{Histogram of estimated rank threshold from real data (January 16 2006 to August 22 2006).
}

\end{figure}

\clearpage
\begin{figure}

\centerline{\includegraphics[width=14cm,bb=0 0 6225 4691]{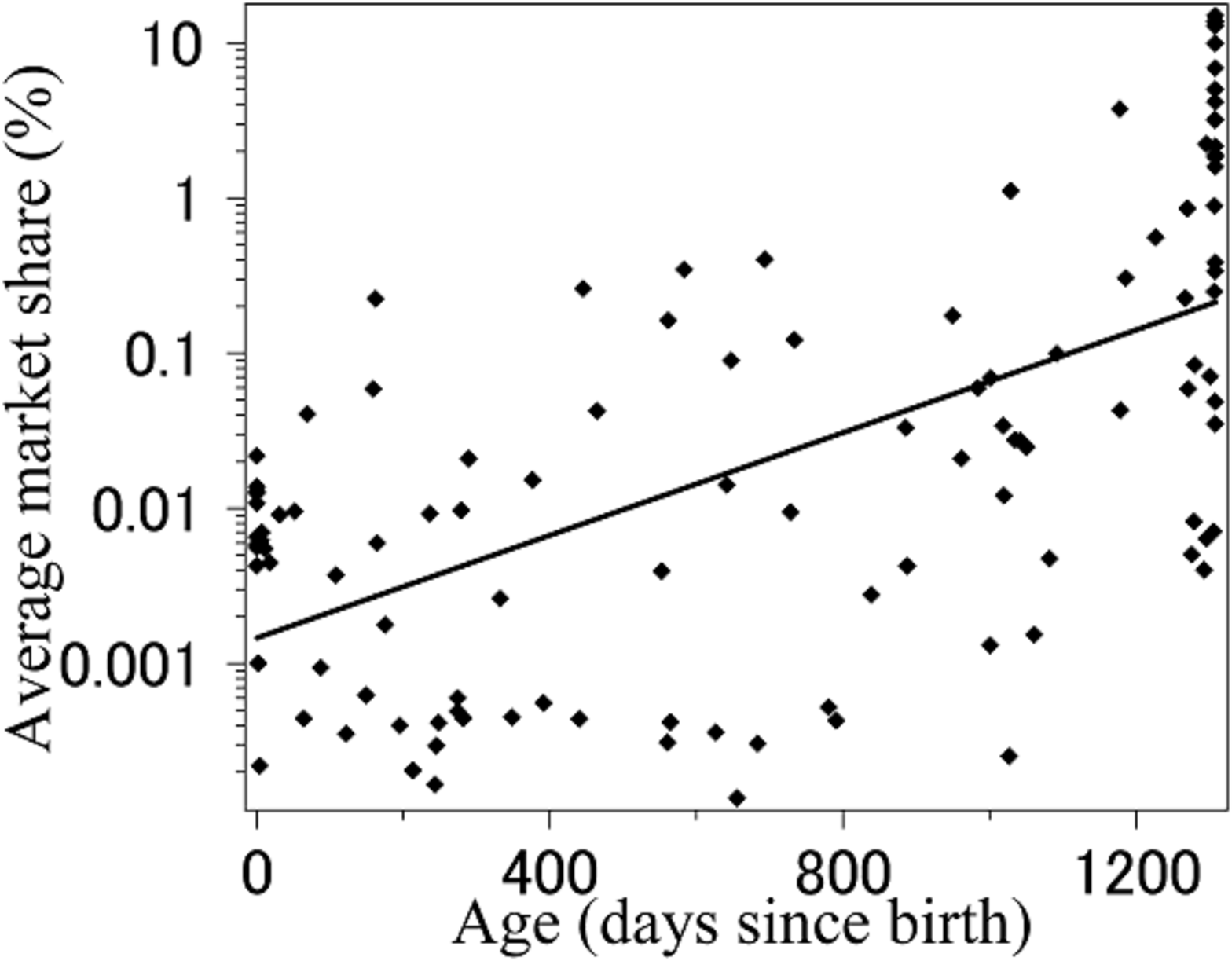}}

\caption{Average daily market share of brands versus its age in semilog scale.  Continuous line shows the least squares fit for the exponential hypothesis.
}

\end{figure}

\clearpage
\begin{figure}

\centerline{\includegraphics[width=14cm,bb=246 159 772 408]{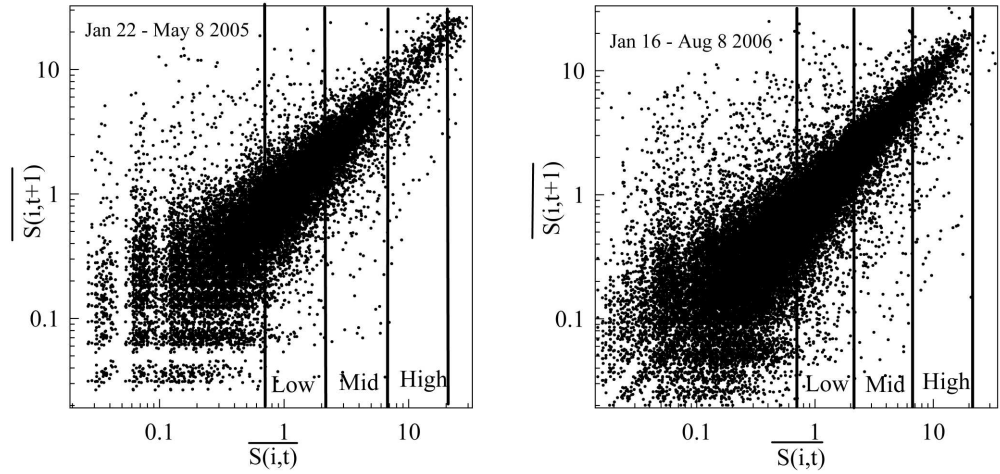}}

\caption{Cutting scatter plots into equal logarithmic bins. Left panel depicts sales of successive weeks from period January 22 2005 to May 8 2005, and right panel depicts sales of successive weeks from period January 16 2006 to August 8 2006.
}

\end{figure}

\clearpage
\begin{figure}

\centerline{\includegraphics[width=19cm,bb=174 251 722 438]{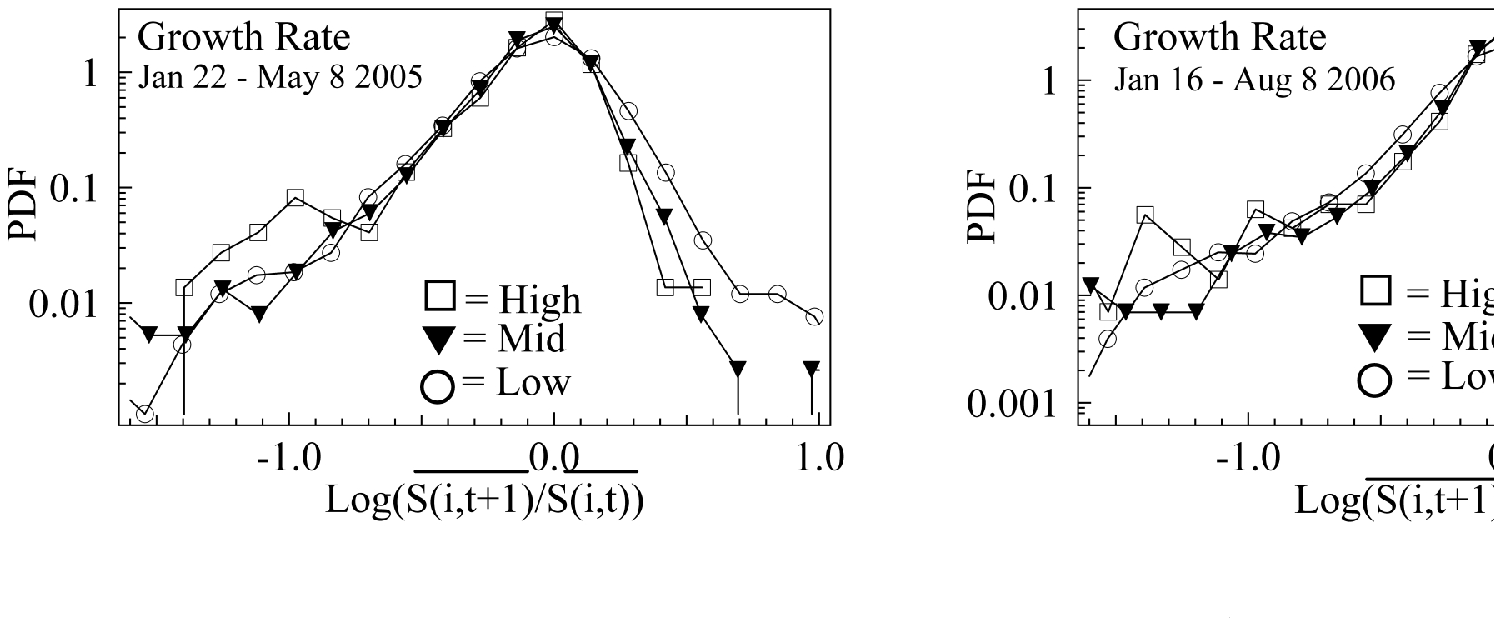}}

\caption{Left panel shows distribution of log growth for one week for period January 22 2005 to May 8 2005. Right panel shows distribution for January 16 2006 to August 8 2006. Although right panel clearly seems size dependent, between the middle and high areas, the left seems to coincide.
}

\end{figure}

\clearpage
\begin{figure}

\centerline{\includegraphics[width=16cm,bb=329 177 672 377]{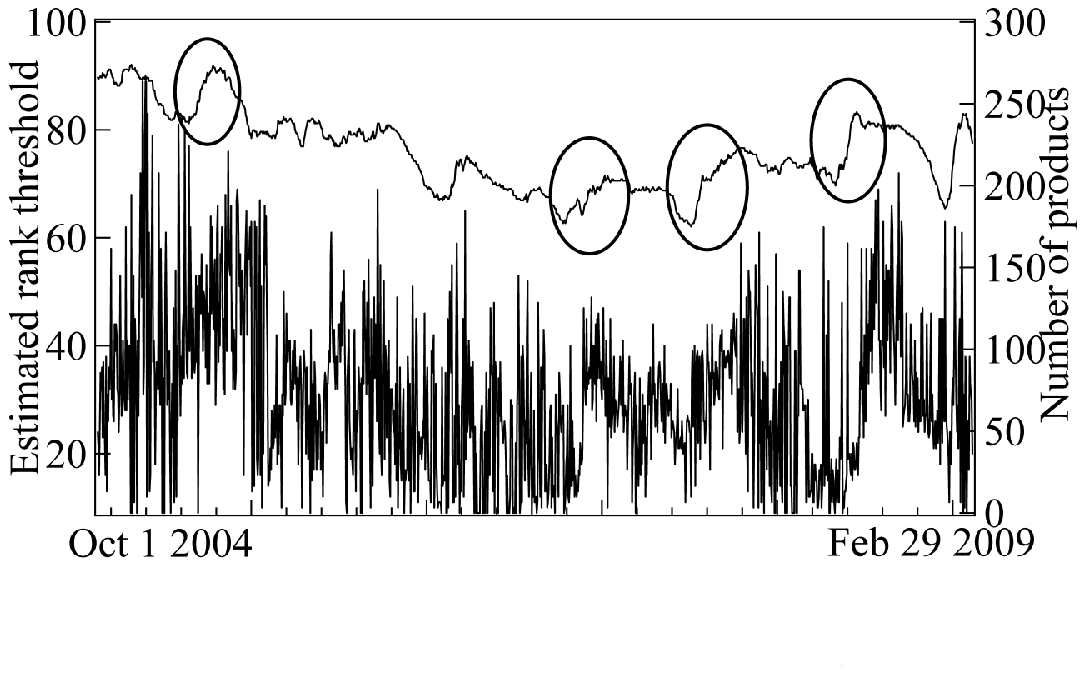}}

\caption{Estimated rank threshold for all dates with time evolution of number of products on market.
}

\end{figure}

\clearpage
\begin{table}

\centerline{\includegraphics[width=16cm,bb=65 420 589 535 ]{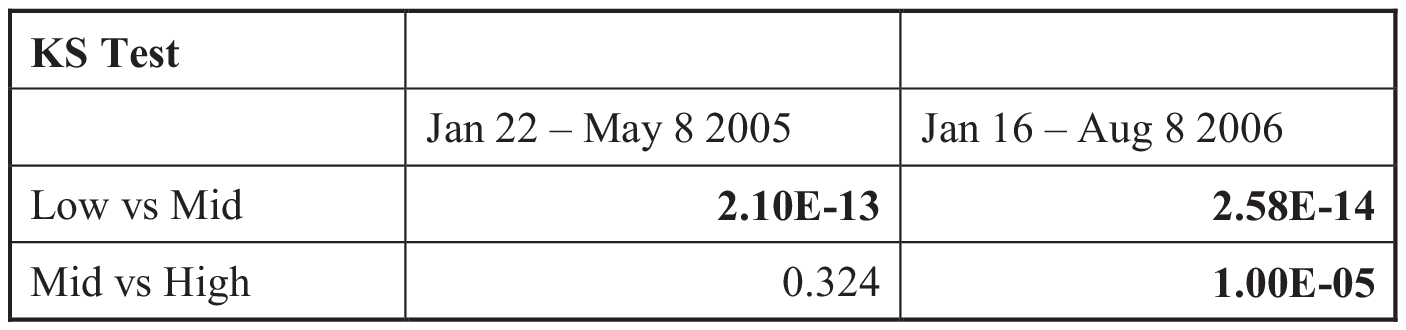}}

\caption{Result of two sample Kormogorv-Smirnov test. Numbers inside show p-values. Rows represents pairs and columns denote period. Null hypothesis states that two distributions are identical, and the alternative states that they are different.
}

\end{table}


\begin{references}
[1] Pareto,V., "Cours D'Economie Politique", Rouge, Lausanne et Paris, (1896).

\\

[2] Mitzenmacher, M., "A Brief History of Generative Models for Power Law and Lognormal Distributions", Internet Mathematics, Vol. 1, No. 2, 226-251, (2004).

\\

[3] Gibrat, R., "Les Inegalites Economiques", Librairie du Recueil Sirey, Paris, (1931).

\\

[4] Auerbach, F., "Das Gesetz der Bvolkerungskonzentrayion. Petermann's Geographiche Mitterilungen", 59, 73-76., (1913).

\\

[5] Estoup. J.B., "Gammes Stenographiques", Paris: Institut Stenographique de France, (1916).

\\

[6] Zipf, G., "Selective Studies and the Principle of Relative Frequency in Language", Cambridge, M.A., Harvard University Press, (1932).

\\

[7] Yule, G., "A Mathematical Theory of Evolution Based on the Conclusions of Dr. J. C. Wills", F.R.S. Philosophical Transactions of the Royal Society of London (Series B), 213, 21-87, (1925).

\\

[8] Lotka, A. J., "The Frequency of Distribution of Scientific Productivity", Journal of the Washington Academy of Sciences, 16, 317-323, (1926).

\\

[9] Willinger, W., and Paxson, V., "Where Mathematics Meets the Internet", Notices of the American Mathematical Society", 45, 961, (1998).

\\

[10] Sornette, D., Deschartes, F., Gilbert, T., and Ageon, Y., "Endogenous Versus Exogenous Shocks in Complex Networks an Empirical Test Using Book Sale Ranking", Physical Review Letters, 93, 228701, (2004).

\\

[11] Koli,R. and Sah, R., "Some Empirical Regularities in Market Shares", Management Science (2006).

\\

[12] Kapteyn, J. and Uven J. M, "Skew Frequency Curves in Biology and Statistics", Hoitsema Brothers, Groningen, (1916). 

\\

[13] Sutton, J., "Gibrat's Legacy", Journal of Economic Literature, Vol. XXXV, pp. 40-59, (1997).

\\

[14] Fu, D., Pammolli, F., Buldyrev, S.V., Riccaboni, M., Matia, K., Yamasaki, K., and Stanley, H.E., "The growth of business firms: Theoretical framework and empirical evidence", PNAS, Dec 27, 2005, Vol. 102, No.52.

\\

[15] Picoli, S., Mendes, R. S., and Malacarne, C., "Statistical properties of the circulation of magazines and newspapers", Europhys. Lett., 72(5), pp. 865-871, (2005).

\\

[16] Ishikawa, A., Fujimoto, S., Mizuno, T., "Shape of Growth Rate Distribution determines the type of Non-Gibrat's Law", arXiv:1003.0940.

\\

[17] http://www.bcn.co.jp/.

\\

[18] Malevergne, Y., Pisarenko, V., and Sornette, D., "Gibrat's law for cities: uniformly most powerful unbiased test of the Pareto against the lognormal", American Economic Review, forthcoming.

\\

[19] Clauset, A., Shalzi, C.R., and Newman, M.E.J, "Power law distributions in empirical data", SIAM Reviews, (2009).

\\

[20] Castillo, J., "The Singly Truncated Normal Distribution, a Non-Step Exponential Family", The Annals of the Institute of Mathematical Statistics, 46, 57-66, (1994).

\\

[21] Castillo, J. and Puig, P, "The Best Test of Exponentiality against Singly Truncated Normal Alternatives", Journal of the American Statistical Association 94, 529-532, (1999).

\\

[22] Klass, O., Biham, O., Levy, M., Malcai, O., and Solomon, S., "The Forbes 400, the Pareto power law and efficient markets", Economics Letters 55, 143-147

\\

[23] Saichev, A., Malevergne, Y., and Sornette, D., "Theory of Zipf's Law and Beyond", Lecture notes in Economics and Mathematical Systems, 632, Springer

\\

[24] Buldyrev, S. V., Pammolli, F., Riccaboni, M., Yamasaki, K., Fu, D-F., Matia, K., and Stanley, H. E., "A generalized preferential attachment model for business firms growth rates II. Mathematical treatment", Eur. Phys. J. B 57, 131-138, (2007).

\\

[25] Sakai, K., and Watanabe, T., "The firm as a bundle of barcodes", Eur. Phys. J. B, DOI: 10.1140/epjb/e2020-00069-6, 2010

\\

[26] Mizuno, T., Takayasu, M., "The Statistical Relationship between Product Life Cycle and Repeat Purchase Behavior in Convenience Stores", Progress of Theoretical Physics Supplement 179, 71-79, 2009.

\\

[27] Takayasu, H., Sato, H., and Takayasu, M., "Stable Infinite Variance Fluctuations in Randomly Amplified Langevine Systems", Physical Review Letters, Vol. 79, No. 6, (1997).

\\

[28] Mizuno, T., Takayasu, M., and Takayasu, H, "The mean-field approximation model of company's income growth", Physica A, 332, 403-411, (2004).
\end{references}
\end{document}